# The Ice Cap Zone: A Unique Habitable Zone for Ocean Worlds

Ramses M. Ramirez[1] and Amit Levi[2]

[1]Earth-Life Science Institute, Tokyo Institute of Technology, 2-12-1, Tokyo, Japan 152-8550
[2] Harvard-Smithsonian Center for Astrophysics, 60 Garden Street, Cambridge, MA 02138, USA

email: rramirez@elsi.jp

**ABSTRACT**
Traditional definitions of the habitable zone assume that habitable planets contain a carbonate-silicate cycle that regulates $CO_2$ between the atmosphere, surface, and the interior. Such theories have been used to cast doubt on the habitability of ocean worlds. However, Levi et al (2017) have recently proposed a mechanism by which $CO_2$ is mobilized between the atmosphere and the interior of an ocean world. At high enough $CO_2$ pressures, sea ice can become enriched in $CO_2$ clathrates and sink after a threshold density is achieved. The presence of subpolar sea ice is of great importance for habitability in ocean worlds. It may moderate the climate and is fundamental in current theories of life formation in diluted environments. Here, we model the Levi et al. mechanism and use latitudinally-dependent non-grey energy balance and single-column radiative-convective climate models and find that this mechanism may be sustained on ocean worlds that rotate at least 3 times faster than the Earth. We calculate the circumstellar region in which this cycle may operate for G-M-stars ($T_{eff}$ = 2,600 – 5,800 K), extending from ~1.23 - 1.65, 0.69 - 0.954, 0.38 – 0.528 AU, 0.219 – 0.308 AU, 0.146 – 0.206 AU, and 0.0428 – 0.0617 AU for G2, K2, M0, M3, M5, and M8 stars, respectively. However, unless planets are very young and not tidally-locked, our mechanism would be unlikely to apply to stars cooler than a ~M3. We predict C/O ratios for our atmospheres (~0.5) that can be verified by the JWST mission.

# 1. INTRODUCTION

In this work, we focus on ocean worlds. These are defined here as: (1) exoplanets with tens of percent of their total mass composed of water that (2) lack a substantial H/He envelope and (3) have migrated close enough to their host star so that their outermost condensed mantle is a global ocean.

Accretion of a few percent by mass of water is sufficient to form a barrier, composed of high-pressure water ice polymorphs, between the rocky inner mantle and the ocean (Levi, Sasselov & Podolak, 2014). A high mass fraction of water may be explained by forming the planet beyond the snowline (Kuchner 2003) (L´eger, et al. 2004). Photo-evaporation of the H/He atmosphere places an upper bound on the planetary core mass of about $2M_\oplus$ (Luger, et al. 2015). However, a low mass planet forming beyond the snowline implies a low mass disk, which is associated with low mass stars. Using population synthesis models, Alibert & Benz (2017) have shown that water-rich planets as defined here are a natural outcome around very low mass (~ M spectral class) stars. The abundance of low mass stars and the ubiquity of water in solar systems suggest our studied planets should be very common, with implications for near future observations. We note that M-class stars experience a particularly large migration in their snow line, likely affecting the formation of water rich planets. This issue continues to be studied, however. Another substantial unknown is the retention of water during high-impact collisions. Plus, recent planet formation models have largely been driven by limited observations. Nevertheless, statistics suggests that ocean worlds may be common throughout the cosmos (Simpson, 2017). Given such uncertainties, our search for the ice cap zone will extend beyond M-class stars (G-class).

A better understanding of the habitable zone (HZ) (which is the circular region around a star where standing bodies of liquid water could be stable on the surface of a rocky planet) is required in order to better guide observations toward planets of hypothesized biological interest. Even though not all HZ planets are habitable, the HZ remains a useful tool for prioritizing promising targets in the search for extraterrestrial life. The most popular incarnation of this concept remains that devised by Kasting et al. (1993) (and updated by us in Kopparapu et al. 2013) which posits that habitable planets have roughly Earth-like masses, Earth-like surface water inventories, a carbonate-silicate cycle, and orbit stars in their main-sequence of stellar evolution. Another key assumption is that $CO_2$ and $H_2O$ are the main greenhouse gases as they are on the Earth, leading to $H_2O$-dominated atmospheres near the inner edge and $CO_2$-dominated atmospheres near the outer edge. In what we dub here as the "classical HZ", the entire surface water inventory evaporates and a runaway greenhouse edge ensues at the inner edge boundary, leading to rapid desiccation and Venus-like planetary conditions. In our solar system, this "runaway greenhouse" limit occurs at ~0.95 AU (Leconte et al., 2013; Kopparapu et al., 2014; Ramirez & Kaltenegger, 2014;2016). A more pessimistic inner edge, the moist greenhouse, occurs when mean surface temperatures exceed ~340 K, beyond which a water inventory equal to the amount in Earth's oceans is lost to space on a timescale of ~4.5 Gyr (i.e. the age of the solar system)(e.g. Kasting et al. 1993). On the other hand, the outer edge of this classical HZ is the distance beyond which the combined effects of $CO_2$ condensation and Rayleigh scattering outweigh its greenhouse effect. The higher $CO_2$ pressures needed to sustain warm surface conditions at distances approaching this outer edge are assumed to be provided by the carbonate-silicate cycle, which helps ensure a relatively wide habitable zone (e.g. Kasting et al. 1993).

Although the classical HZ is a great way to frame planetary habitability, how well this particular formulation represents reality is, at present, unknown and requires continued study. An ongoing discussion regarding how representative the underlying assumptions may be for the habitability of extraterrestrial planets continues (e.g. Abe et al., 2011; Agol, 2011; Pierrehumbert & Gaidos, 2011; Seager, 2013; Zsom et al. 2013; Kasting et al. 2014; Ramirez & Kaltenegger, 2014; 2016;2017).

In here, we revisit the notion of the carbonate-silicate cycle, which regulates the exchange of $CO_2$ between the interior and the atmosphere and is thought to be vital in maintaining clement surface temperatures over Earth's geologic history (e.g. Kasting et al. 1993). Abbot et al. (2012) have argued that such a cycle can be sustained on worlds with land fractions as low as 5%. Although this suggests that the cycle would operate on planets that have at least some land, it also indicates that true ocean worlds may be uninhabitable.

However, studies suggest this may not be the case. In the absence of a silicate weathering feedback, atmospheric $CO_2$ concentrations are expected to build up in ocean worlds (e.g. Wordsworth and Pierrehumbert, 2013). Moreover, the resultant $CO_2$ cycle would exhibit a stabilizing negative feedback with high $CO_2$ cooling rates favoring lower atmospheric escape rates (e.g. Wordsworth and Pierrehumbert, 2013; Kitzmann et al., 2015).

Clearly then, the classical HZ is inadequate for describing our studied ocean worlds. A planet much richer in water than the Earth (tens of percent of the total planetary mass) will not likely lose its surface water inventory in a runaway greenhouse. In addition, though the carbonate-silicate cycle is inactive due to the high-pressure ice barrier, other mechanisms, based on water ice phase transitions in sea ice, may moderate the climate (Levi et al. 2017). Thus, the difference in the geophysical cycles between rocky planets and ocean worlds merit an independent definition for the inner and outer edges of the HZ for the case of ocean worlds.

As explained above, the classical HZ does not explicitly consider needed conditions for the evolution of life, besides the stable existence of liquid water. In this work, we attempt to incorporate within our definition of the HZ for ocean worlds restrictions imposed by prebiotic chemistry and the initiation of an RNA-world. This, we expect, should yield a more restrictive, yet stronger, filter for targeting planets for observation.

A major concern when considering the evolution of life on ocean worlds is that the absence of continents and the deep ocean yield a globally diluted environment. For example, formaldehyde and HCN are important precursors in several synthetic pathways, known to produce amino acids and nucleobases, see discussion in Schwartz and Goverde (1982). Reactions forming amino acids and nucleobases require a high concentration of HCN, otherwise hydrolysis of HCN dominates, resulting in the production of formate and ammonia.

Levy et al. (2000) suggested that when a diluted aqueous solution freezes, the ice Ih grains exsolve impurities that become concentrated within the remaining liquid pores. The high concentrations obtained more than offset the lower kinetics at subfreezing temperatures. The high concentrations thus achieved can drive production at high yields. Freezing a 0.1M solution of $NH_4CN$ produces: glycine, racemic alanine and aspartic acid (Levy, et al. 2000). Sparking an aqueous solution of $CH_4$, $N_2$ and $NH_3$ formed adenine, only upon freezing to $-20°C$ (Levy, et al. 2000). Kanavarioti et al. (2001) investigated ice as a reaction medium using uridine nucleotides. They found that freezing to $-18°C$ was a necessary condition for enhanced oligomerization, with yields decreasing for higher temperatures. Therefore, frozen conditions are superior to free liquid solutions in synthesizing RNA oligomers. Furthermore, freezing was found to support and enhance RNA replication without comprising fidelity, and provide a form of compartmentalization required by Darwinian evolution (Attwater, et al. 2010).

However, others suggest that mere freezing is not sufficient. Menor-Salvan et al. (2009) found that spark discharges in a reduced atmosphere highly favored the production of cytosine and triazine, only when the system was exposed to a freeze-thaw cycle (between $-5°C$ and $5°C$). In the absence of a freeze-thaw cycle the production of tholins was favored. Freeze-thaw cycles are also needed to help bridge the gap from the prebiotic chemistry of short RNA oligomers to complex replicating ribozymes (Mustschler, Wochner & Holliger, 2015).

In ocean worlds freeze-thaw cycles are possible where there is a polar ice cap. Forming sea ice ought to gradually exsolve impurities into its pores, approaching eutectic compositions. Thus, providing the necessary increase in concentrations discussed above. Further transition of the ice Ih grains into $CO_2$ clathrate hydrate comes at the expense of pore space, hence increasing the chances of molecular collisions and reactions. Some sea ice floes on the periphery of the subpolar region may dissociate, due to migration to latitudes with higher surface temperatures, rather than sink (see discussion in Levi et al., 2017). It is possible that processed material released from the pores, following this dissociation, will refreeze, if returned by surface currents to higher latitudes.

Hence, pinpointing the heliocentric region around different types of stars, which is coincident with the condition of having a subpolar ice cap, is important. We refer to this region as the ice cap zone. Planets closer in are free of sea ice and planets farther out are snowballs.

Although the work in Levi et al. (2017) suggests that ocean worlds may be habitable without the carbonate-silicate cycle, the ocean model was not coupled to an atmospheric model to assess the plausibility of such solutions. Here, we couple the model in Levi et al. (2017) with a single-column radiative-convective climate model and an energy balance climate model developed by one of us (Ramirez) to determine the location of the ice cap zone for G- M-stars. In Section 2 we give a brief description of the $CO_2$ ocean-atmospheric exchange modelled in Levi et al. (2017) and describe the inner and outer boundaries of the ice cap zone from a geophysical perspective. The climatological perspective is given in Section 3.1. In Sections 3.2 and 3.3 we describe the radiative-convective and energy balance climate models used to perform our analyses. Climate modeling procedures are discussed in section 3.4. Our Results and Discussion are found in Sections 4 and 5, respectively, leading to the Conclusion.

## 2. THE ICE CAP ZONE

For the benefit of the reader the following is a brief review of the mechanism developed in Levi et al. (2017), and a recap of a few of its fundamental governing equations. In Levi et al. (2017) we have studied different mechanisms governing the exchange of $CO_2$ between the deep water-rich ice mantle, the ocean, and the atmosphere (see Fig.1 for an illustration of our system). We have suggested that $CO_2$ in the water-rich ice mantle promotes an ocean floor enriched in $CO_2$ clathrate hydrates. An ocean under-saturated in $CO_2$ would cause the dissociation of these clathrates, and the subsequent release of $CO_2$ into the ocean, thus reaching saturation. Therefore, the clathrate phase controls the $CO_2$ saturation level in the deep (~100km) global ocean.

Vertical mixing and homogenization of the oceans here on Earth are powered by tides and winds. These supply about 1TW, which is sufficient to support the circulation in Earth's oceans (Wunsch, C. & Ferrari, R. 2004; Kuhlbrodt et al. 2007). However, this power is insufficient for homogenizing the deep oceans that are the focus of this study, requiring at least tens of TW (Levi et al. 2017). Winds, however, can support the circulation of the upper ocean (upper few kilometers), with implications for $CO_2$ outgassing rates.

A meridional temperature gradient, on a rotating planet, should give rise to surface wind patterns that cause the ocean surface water to diverge and converge. This promotes circulation in the upper ocean via Ekman pump and suction (see illustration in Fig.1). The net outgassing of dissolved $CO_2$ into the atmosphere, due to this circulation, is the difference between the influx and outflux of $CO_2$ between ocean and atmosphere.

In regions where oceanic surface water converge, Ekman pumping results in an influx of atmospheric $CO_2$ of (see eq.53 in Levi et al. 2017):

$$j_{CO_2}^{in} = [\tilde{\beta}(T_{sbt}) - \tilde{\beta}(T_{trop})]P_{atm}^{CO_2} w_e \quad (1)$$

Here $\tilde{\beta}$ is a Henry-like constant relating solubility (dissolved number density of $CO_2$) to the partial atmospheric pressure of carbon-dioxide, $P_{atm}^{CO_2}$. The vertical velocity in the circulation is $w_e$. The subtropical and tropical oceanic surface temperatures are: $T_{sbt}$, and $T_{trop}$, respectively.

At the bottom of the wind-driven circulation, the geostrophic flow is in contact with the deep ocean, which is saturated in $CO_2$. Therefore, enriching this water with $CO_2$, if it is under-saturated. The liquid parcels that are reemerging at the ocean surface, in regions of surface water divergence, experience degassing, thus establishing a $CO_2$ outflux of (see eq.76 in Levi et al. 2017):

$$j_{CO_2}^{out} = w_e[n_{CO_2}^{out} - \tilde{\beta}(T_{trop})P_{atm}^{CO_2}] \quad (2)$$

where, $n_{CO_2}^{out}$ is the number density of $CO_2$ dissolved in the enriched water pumped to the surface via Ekman suction. The latter is a complex function of time, the vertical eddy diffusion coefficient for the deep ocean, the number density of $CO_2$ dissolved in the deep ocean, the ocean depth, the flux of circulated water that comes in contact with the deep unmixed ocean, and the length scale between regions of surface water convergence and divergence. These relations are given in Eqs.71-75 in Levi et al. (2017)

The fluxes due to the wind-driven circulation equilibrate when the number density of $CO_2$ dissolved in surface regions where the ocean water converges, equals the number density of dissolved $CO_2$ in the saturated deep ocean. This results in an atmosphere of tens of bars of $CO_2$ (see Fig.20 in Levi et al. 2017). However, such an atmosphere is unstable if anywhere on the surface of the planet temperatures are lower than about 280K. In these colder regions, atmospheric $CO_2$, at the level of a few tens of bars, will convert the liquid surface water into $CO_2$ clathrate hydrate, on the expense of the atmospheric $CO_2$. The clathrate hydrate of $CO_2$ is denser than the ocean (see Fig.23 in Levi et al.2017), and is thermodynamically stable in the deep saturated ocean. Therefore, atmospheric $CO_2$ deposited in these clathrates will sediment to the bottom of the ocean, thus, lowering the atmospheric pressure of $CO_2$ toward its clathrate dissociation pressure (a few bars). Therefore, a constant tension exists between two tendencies: on the one hand trying to equilibrate the atmospheric abundance of $CO_2$ with that which is dissolved in the deep ocean, while on the other hand sinking any excess in the atmospheric $CO_2$ above the clathrate dissociation pressure.

For subpolar temperatures below 250K water first freezes as ice Ih, which is then converted over time into a $CO_2$ clathrate hydrate, as explained above. When a mole fraction, $\alpha_{min}$, of the original ice ih is transformed into $CO_2$ clathrate hydrate, the sea-ice composite becomes denser than the ocean and sinks (see eq. 89 in Levi et al. 2017). The time elapsed before a clathrate enrichment of $\alpha_{min}$ is reached is a function of the temperature, pressure and ice grain morphology. The resulting number of $CO_2$ molecules removed from the atmosphere, and deposited on the bottom of the ocean, per unit time is (see eq. 94 in Levi et al. 2017):

$$\frac{dN_{ice-sink}}{dt} = \frac{2\pi R_p^2(1-sin\lambda_{sp})h_{ice}\varsigma\alpha_{min}(1-\phi_{pore}^0)}{\Delta\tau} \frac{46}{V_{cell}} \frac{1}{5.75} \quad (3)$$

Here $\lambda_{sp}$ is the latitudinal extent of the ice cap, $R_p$ is the planetary radius, $h_{ice}$ is the thickness of the sea-ice slab prior to sinking, $\phi_{pore}^0$ is the initial porosity of the forming sea-ice, $V_{cell}$ is the unit cell volume of the clathrate, and $\varsigma$ is an expansion coefficient equal to 1.133. $\Delta\tau$ is the time duration a sea-ice slab remains afloat, i.e. the time needed to reach a clathrate enrichment of $\alpha_{min}$. $\Delta\tau$ is restricted by the divergent flow of sea-ice away from the cold subpolar region (see eq.103 in Levi et al. 2017). In other words, sea-ice remaining positively buoyant beyond this time criterion, may be carried away to warmer regions, where its clathrate component may dissociate, consequently releasing its $CO_2$ content back into the atmosphere. This restriction on $\Delta\tau$ leads to a restriction on the ice grain size (We refer the interested reader to subsection 6.2 in Levi et al. 2017, for more details and an in-depth discussion of this issue).

Considering the fluxes above, a steady state for the partial atmospheric pressure of $CO_2$ is reached over time, which obeys the following relation (see eq. 102 in Levi et al. 2017):

$$0 = \pi R_p N_{wdc}\tilde{\beta}(T_{sbt})D_{eddy}\frac{L_g}{L_{ocean}}\left[\frac{n_{CO_2}^{deep}}{\tilde{\beta}(T_{sbt})} - P_{atm}^{CO_2}\right] - S_w\frac{dN_{ice-sink}}{dt} - Q_c\frac{4\pi R_p^2}{m_{CO_2}g}$$
(4)

The three terms on the right-hand side of the last equation represent the contributions of: the wind-driven circulation (i.e. equilibration with the deep ocean), the sea-ice sink mechanism, and an external atmospheric erosion term, respectively. $N_{wdc}$ is the number of wind driven circulations, $D_{eddy}$ is the vertical eddy diffusion coefficient for the deep ocean, $L_g$ is the horizontal length scale of the circulation, $L_{ocean}$ is the depth of the ocean, $n_{CO_2}^{deep}$ is the number density of dissolved $CO_2$ in the deep ocean, $m_{CO_2}$ is the mass of a $CO_2$ molecule, $g$ is the acceleration of gravity, and $Q_c$ is an external atmospheric erosion factor having units of pressure over time.

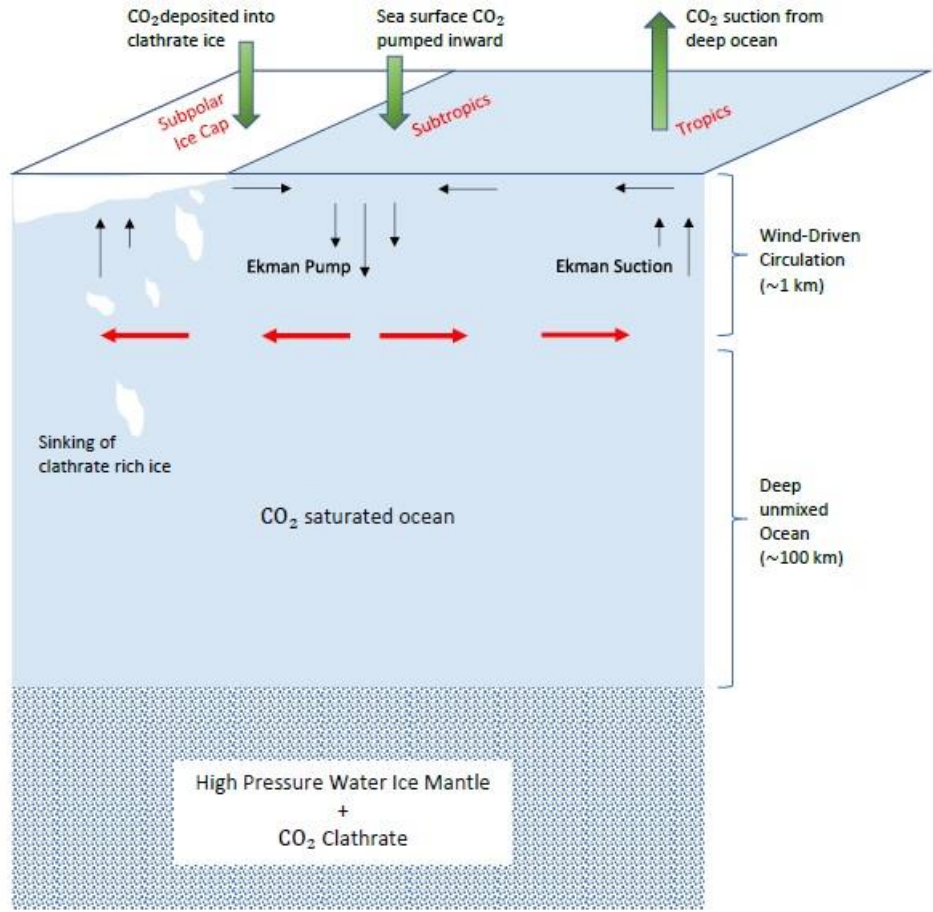

**Figure 1:** An illustration of the transport of $CO_2$ between ocean and atmosphere in water-rich planets. The ocean floor is rich in $CO_2$ clathrate hydrate, which keeps the ocean saturated in $CO_2$. Winds drive a circulation of the upper ocean. The circulated water is enriched in $CO_2$ due to the contact with the deep saturated ocean. This added $CO_2$ is later degassed into the atmosphere during the suction of deep water. This process strives to outgas tens of bars of $CO_2$ into the atmosphere. However, this high atmospheric pressure may be unstable. If subpolar temperatures are subfreezing, then sea-ice is transformed over time from mainly being composed of ice Ih into a composition rich in the clathrate hydrate of $CO_2$. A sufficient enrichment in the latter phase turns the sea-ice floe negatively buoyant. This process tends to clear the atmosphere of any excess $CO_2$, thus reducing the global atmospheric pressure of $CO_2$ to the dissociation pressure of the clathrate phase.

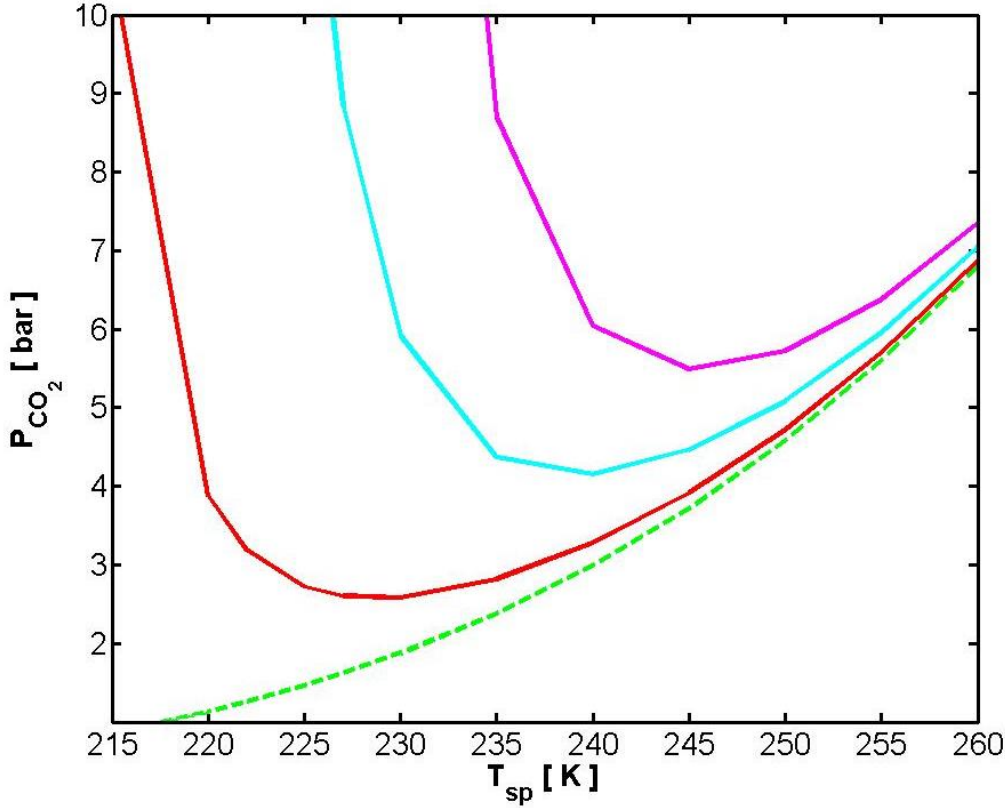

**Figure 2:** The steady-state partial atmospheric pressure of $CO_2$ as a function of the average polar surface temperature. The vertical axis is the steady-state atmospheric $CO_2$ pressure. The horizontal axis is the subpolar temperature averaged over the planetary orbit. The following initial ice Ih grain radii are assumed for the sea ice: 100μm (solid red curve), 200μm (solid cyan curve) and 300μm (solid magenta curve). The dashed green curve represents the dissociation pressure of the $CO_2$ clathrate hydrate. The latter phase is stable only above its dissociation curve.

In Fig. 2 (see also Fig.29 in Levi et al. 2017) we plot solutions for the steady-state partial atmospheric pressure of $CO_2$ versus the subpolar temperature. The rate of transformation from ice Ih into clathrate is temperature- and pressure-dependent. This transformation is kinetically fast for the highest temperatures plotted in Fig. 2. Therefore, the resulting steady-state atmospheric pressure is close to the dissociation curve of the clathrate (green dashed curve). This is because the sea-ice sink mechanism is effective. When considering lower temperatures, for the subpolar region, the transition between the ice phases slows down. As a result, sea-ice can migrate away from the ice cap region to a warmer climate, and release its $CO_2$ content back into the atmosphere. The increase in $CO_2$ pressure acts to accelerate the transition between the phases, counteracting the effect of the lower temperature. Clearly, the lower the temperature is, the higher is the $CO_2$ pressure needed to counterbalance the effect of the former. Thus, a minimum in the pressure is evident for some subpolar temperature, depending on the morphology of the ice grains forming the sea-ice (see Levi et al. 2017 for a discussion on the ice grain morphology).

In Levi et al. (2017) we have estimated that it takes thousands of years for the wind-driven circulation to smooth out disturbances in the atmospheric pressure from its steady state value. Near steady state the sea ice sink for atmospheric $CO_2$ can change the atmospheric abundance of this chemical species at a rate of about 2 mbar $yr^{-1}$. This is assuming the polar ice cap extends to the $60^{th}$ parallel and an initial sea ice porosity of 0.1. The possible change in atmospheric pressure per year is much less than the atmospheric pressure predicted at steady state. Therefore, the influxes and outfluxes of $CO_2$ are far too low for the atmosphere to adjust to seasonal variations in the subpolar temperature. Hence, in this work, the geophysical model is coupled with the climate model using a subpolar temperature which is averaged over the planetary orbit.

The ice cap zone has an inner and outer edge. In this section, we wish to address these edges within a geophysical context. We discuss this issue further within a climatological context below (See Section 3.1).

Planets at stellar distances where (eqn. 5),

$$\frac{dP_{CO_2}}{dT_{sp}} > 0 \qquad (5)$$

and that experience an increase in irradiation, will have a heating ice cap and further buildup of $CO_2$ in their atmosphere, increasing the greenhouse effect. This may lead to a runaway effect, transitioning these planets to a state free of sea ice (and a buildup of tens of bars of $CO_2$ in their atmospheres). However, a decrease in irradiation will cool the ice cap, causing the sea ice sink mechanism to further reduce the atmospheric pressure of $CO_2$. Thus, driving the surface condition toward those corresponding to the minimum point in Fig.2.

Planets at stellar distances where (eqn. 6),

$$\frac{dP_{CO_2}}{dT_{sp}} < 0 \qquad (6)$$

and that experience an increase in irradiation, will have a heating ice cap, though now $CO_2$ will sink out of the atmosphere decreasing the greenhouse effect. This implies re-cooling of the ice cap. A decrease in irradiation, will cool the ice cap, causing further build-up in atmospheric $CO_2$. This implies re-heating of the ice cap. Considering that only where the gradient is negative an ice cap may be stable over a long period of time, we suggest that the condition (eqn. 7),

$$\frac{dP_{CO_2}}{dT_{sp}} = 0 \qquad (7)$$

demarcates the inner edge of the ice cap zone.

The outer edge of the ice cap zone is related to the ice-albedo feedback mechanism, which is discussed below (Section 3). Here we estimate the activation temperature of this mechanism, as influenced by the introduction of $CO_2$ clathrate hydrates into the forming sea ice. The albedo of sea ice depends on various factors discussed below. Here we stress its dependence on the thickness of the sea ice floe. The albedo is found to increase substantially in the thickness range of tens of centimeters (Brandt, et al. 2005) (Allison et al., 1993) (Perovich 1990).

On a planet with an Earth-like atmospheric composition, sea ice is mostly composed of ice Ih. This phase is less dense than the ocean water, making sea ice gravitationally stable. A water-rich planet likely has a dense $CO_2$ atmosphere. Therefore, sea ice would experience a compositional transition over time, from a predominantly ice Ih structure to one enriched in $CO_2$ clathrate hydrates. In Levi et al. (2017) it was shown that, when sea ice enrichment in the latter phase reaches a molar fraction of approximately 0.4, the sea ice composite becomes gravitationally unstable. Consequently, the sea ice sinks to the bottom of the ocean, exposing open water, with its associated lower albedo, to the planetary surface.

The ice-albedo feedback mechanism requires a stable ice cover. This condition is satisfied, in the case of our studied planets, when the time scale for the formation of a thick sea ice layer is less than the time scale for it becoming gravitationally unstable. The rate of thickening of sea ice is controlled by the need to conduct the latent heat of fusion, released at its bottom surface, as it forms. Therefore, the time scale of formation of an ice layer of thickness $h$ is (eqn. 8,

$$\tau \approx \frac{L_f \rho h^2}{2\kappa \Delta T} \qquad (8)$$

where $\rho$ is the mass density and $L_f$ is the latent heat of fusion of ice Ih taken from (Feistel and Wagner 2006). The thermal conductivity of ice Ih, $\kappa$, is adopted from (Slack 1980). The temperature difference across the ice layer is $\Delta T = T_m - T_s$, where $T_m$ is the melting temperature of ice Ih, and $T_s$ is the subfreezing surface atmospheric temperature. The rate of conversion of ice Ih to $CO_2$ clathrate hydrate was studied in (Genov, et al. 2004). This rate gives the time it takes the sea ice composite to reach gravitational instability and sink. This time scale also depends on $\Delta T$.

In Fig. 3 we compare between the time scales. In order to be consistent with the values adopted for the ice albedo in the climatological model (see Section 3.4) we solve for two sea ice thicknesses: 30cm (upper limit on young ice) and 50cm (thin first-year sea ice). A $\Delta T$ larger than 5.5K is needed in order to stabilize a 30cm thick sea ice layer. For the thicker ice layer case, a $\Delta T$ larger than 9.1K is required. Taking the average, we conclude, that the surface temperature everywhere needs to drop below about 265K (i.e. $\Delta T = 7.3K$) for there to be a stable and global surface cover of ice.

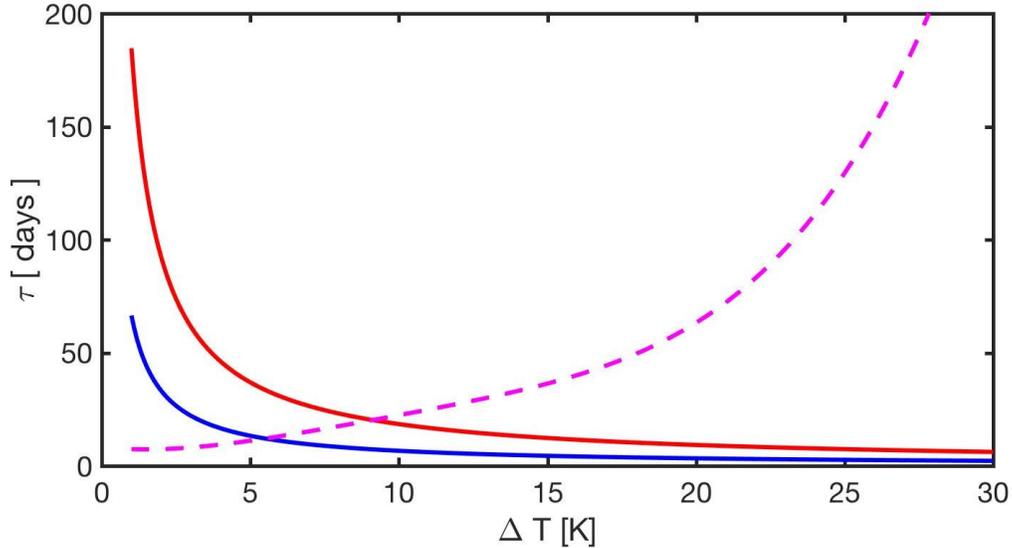

**Figure 3**: Comparison of formation time scales for the formation of 30cm thick sea ice (solid blue curve) and 50cm thick sea ice (solid red curve) and the time before the sea ice becomes gravitationally unstable (dashed magenta curve). $\Delta T$ is the degree of surface temperature subfreezing (see text for definition). For a low degree of subfreezing the rate of sea ice formation is low in comparison to the rate of transformation of the sea ice into a clathrate-rich composite, followed by its sinking into the deep ocean. At a high degree of subfreezing the ice growth rate is high enough to establish a stable surface ice layer, of the appropriate thickness, and maintain a high surface albedo.

## 3. METHODS

*3.1 The Climatological Definition of the Ice Cap Zone*

As with the classical HZ (e.g. Kasting et al. 1993), it is possible to define a circumstellar region for habitable ocean worlds in which $CO_2$ pressures are minimized at the inner edge and maximized at the outer edge. However, our restricted ice cap zone imposes the additional constraints that 1) subpolar sea ice is present for the formation of $CO_2$ clathrate hydrate, and that 2) subtropical temperatures are warm enough to avoid global glaciation, maintaining the wind-driven circulation. Given these conditions, we define our inner edge as the closest distance to the star such that the average subpolar (we define as latitudes between 60 – 90°) surface temperature is a) equal to a value that is determined by the sea ice (Ih) grain size and that b) mean surface subtropical (0 – 60° latitudes) temperatures remain above 265 K. Since sea ice that is rich in $CO_2$ clathrate is denser than the ocean and sinks once a thickness of ~ 1m is achieved (Levi et al. 2017), it is more difficult to glaciate our worlds than the Earth. Indeed, temperatures below the freezing point of water are required (265 K in our model). For the former assumption, we evaluated this limit for 3 representative grain sizes (100, 200, 300 microns). At a grain size of 100 (200, 300) microns, inner edge $CO_2$ pressures of 2.58 (4.15, 5.5) bars, respectively, were required to support subpolar temperatures of 230 (240, 245) K, respectively.

Unlike the inner edge, the outer edge is insensitive to ice Ih grain size. The definition of the outer edge is relatively straightforward as it is the distance beyond which equatorial temperatures (~-5 to +5 degrees latitude) can remain above 265 K, which leads to global glaciation. The $CO_2$ pressure at which this global glaciation ensues depends on the star type (e.g. Kasting et al., 1993).

*3.2 Single-Column Radiative-Convective Climate Model*

We use a single-column radiative-convective climate model first developed by Kasting et al. (1993) and updated most recently (Ramirez & Kaltenegger, 2016; 2017; Ramirez, 2017). The standard model atmosphere is subdivided into 100 vertical logarithmically-spaced layers that extend from the ground to the top of the atmosphere. Absorbed and emitted radiative fluxes in the stratosphere are assumed to be balanced. Should tropospheric radiative lapse rates exceed their moist adiabatic values, the model relaxes to a moist $H_2O$ adiabat at high temperatures, or to a moist $CO_2$ adiabat when it is cold enough tor $CO_2$ to condense (Kasting, 1991).

The model employs 8-term HITRAN and HITEMP coefficients for $CO_2$ and $H_2O$, respectively, truncated at 500 $cm^{-1}$ and 25 $cm^{-1}$, respectively, computed over 8 temperatures (100, 150, 200, 250, 300, 350, 400, 600 K), and 8 pressures ($10^5$ – 100 bar) (Kopparapu et al. 2013; Ramirez et al. 2014ab).

Far-wing absorption in the 15-micron band of $CO_2$ utilizes the 4.3-micron region as a proxy (Perrin and Hartman, 1989). In analogous fashion, the BPS water continuum of Paynter and Ramaswamy (2011) is overlain over its region of validity (0 - ~18,0000 $cm^{-1}$). Also, $CO_2$-$CO_2$ CIA (Gruszka & Borysow, 1997; Gruszka & Borysow, 1998; Baranov et al., 2004; Wordsworth et al. 2010) and $N_2$ foreign-broadening are all implemented (see Ramirez et al. 2014; Ramirez, 2017 for more details). Bt-Settl (Allard 2003; 2007) modeled stellar spectra are used for stars of effective stellar temperatures ranging from 2,600 to 10,000 K (e.g. Ramirez & Kaltenegger, 2016; 2017).

*3.3 Energy balance model Description*

Our energy balance model (EBM) is a state-of-the art non-grey latitudinally-dependent model that is similar to that described in detail by North & Coakley (1979), Williams and Kasting (1997), Caldeira & Kasting (1992), Batalha et al. (2016), Vladilo et al. (2013;2015), Forgan (2016),and Haqq-Misra et al. (2016). These models are well-suited to explore parameter space because they are computationally cheap but are complex enough to produce realistic solutions. In contrast, global circulation models (GCMs) have even more detailed physics but are more difficult to use for such parameter space exploration because of their relatively enormous computational expense. Also, such complex models work best when values for the various planetary parameters they require (e.g. atmospheric composition, land/ocean fraction, rotation rate, relative humidity, topography etc.) are well-known and observable, as they are for the Earth. Thus, because we don't know what most of these quantities for given exoplanets are (which ultimately requires making various assumptions that can't be self-consistently calculated), we believe that simpler models, like ours, are sufficient exploratory tools for the present analysis. Similar opinions as ours have been expressed by others (e.g.

Jenkins, 1993; Vladilo et al., 2015). Moreover, the ocean world atmospheres we assess in this study are dense enough that we can assume efficient planetary heat transfer, even under tidally-locked conditions (e.g. Haberle et al., 1996). Our ocean world planets also do not suffer dynamical complications arising from coupling continental and ocean heat transport (e.g. Cullum et al. 2014; see Section 5 below as well), which also makes an EBM a good analysis tool for this problem. Nevertheless, we present an initial set of results that can be checked against both simple and more complex models in the future.

As with other EBMs, our model's operating principle is that planets in thermal equilibrium must (on average) radiate as much outgoing radiation to space as they absorb from their stars. This version of the model divides the planet into 18 latitudinal zones, each 10 degrees wide. The radiative and dynamic energy balance for each zone is expressed by eqn. 9 (North and Coakley, 1979):

$$C \frac{\partial T(x,t)}{\partial t} - \frac{\partial}{\partial x} D(1-x^2) \frac{\partial T(x,t)}{\partial x} + I = S(1-A) \quad (9)$$

Here, x is the sine of latitude, T is the zonally-averaged surface temperature, S is the incident solar flux, A is the top-of-atmosphere albedo, I is the outgoing infrared flux, C is the effective heat capacity (ocean plus atmosphere), and D is the diffusion coefficient. A second order finite differencing scheme is used to solve the above expression.

Whereas the 1-D radiative-convective climate model provides mean *vertically-*averaged quantities, including pressure-temperature, mixing ratio, and flux profiles, the EBM provides a second dimension in *latitude*-space. Unlike many EBMs, however, which assume grey radiative transfer, we parameterize the outgoing radiation, OLR ($pCO_2$, T), planetary albedo, A ($pCO_2$,T,z, $a_s$), and stratospheric temperature, $T_{strat}$($pCO_2$,T,z) from our 1-D radiative-convective climate modeling results (e.g. Ramirez et al., 2014; Ramirez & Kaltenegger, 2014;2016;2017). Here, $pCO_2$ is the partial pressure of $CO_2$, $a_s$ is surface albedo, and z is the zenith angle. Other non-grey EBMs have parameterized these quantities using complex polynomial fits (e.g. Williams & Kasting, 1997; Haqq-Misra et al. 2016), which can be inaccurate, accruing large errors (up to ~20%). Instead, our EBM achieves >99% accuracy in these quantities through a fine-grid interpolation of OLR, A, and, $T_{strat}$ over a parameter space spanning $10^{-5}$ bar < $pCO_2$ < 35 bar, 150 K < T < 390 K, and 0 < $a_s$ < 1 across all zenith angles (0 to 90 degrees).

The model distinguishes between land, ocean, cloud, and ice cover. Following Haqq-Misra et al. (2016), the model assumes a stellar-dependence on the near-infrared and visible absorption qualities of ice (Shields et al. 2013). $CO_2$ ice is parameterized according to Haqq-Misra et al., (2016). We have implemented a GCM temperature parameterization for the albedo of snow/ice mixtures (Curry et al., 2001) and have included a modified version of the ice-albedo feedback mechanism of Fairén et al. (2012). Instead of idealized Fresnel reflectance data (Kondrat'ev, 1969), which gives big errors at large incidence angles (Briegleb et al., 1986), we used an empirically-determined ocean albedo parameterization as a function of incidence angle for Earth-like planets (ibid).

The current model assumes that heat is transferred via diffusion, and as in previous EBMs (e.g. Williams & Kasting,

1997), this diffusion parameterization has no latitudinal dependence (eqn. 10):

$$\frac{D}{D_o} = \left(\frac{p}{p_o}\right)\left(\frac{c_p}{c_{p,o}}\right)\left(\frac{m_o}{m}\right)^2\left(\frac{\Omega_o}{\Omega}\right)^2 \quad (10)$$

Where $p$ is the pressure, $c_p$ is the heat capacity, $m$ is the atmospheric molecular mass, and $\Omega$ is the rotation rate. Parameters with the subscript "$o$" refer to terrestrial values.

All atmospheres are assumed to be fully-saturated and contain 1 bar $N_2$, with varying amounts of $CO_2$, following the atmospheric composition for the HZ as originally defined by Kasting et al. (1993). Sensitivity studies reveal that decreasing the tropospheric relative humidity to 50% at the outer edge decreases the distance by only ~1 - 2% (although dry atmospheres would yield much bigger ~>10% differences). The outer edge distance is relatively resistant to relative humidity assumptions because associated water vapor concentrations are low (e.g. Godolt, 2016).

The inverse dependence on rotation rate suggests that as the planet's rotation rate is more rapid, the Coriolis effect should inhibit latitudinal heat transport, in agreement with theoretical considerations (e.g. Farrell, 1990). Moreover, the parameterization in eqn. 6 implicitly includes dynamical effects from eddy circulations (Williams and Kasting, 1997; Gierasch and Toon, 1973). Given that our mechanism requires large equator-to-pole temperature gradients to have a stable ice cap and open ocean at the tropics (Levi et al. 2017), increasing $\Omega$ maximizes these temperature gradients by decreasing D.

We vary $\Omega$ from $1\Omega_o$ - $10\Omega_o$, with the latter corresponding to the fastest rotation rate possible for a terrestrial planet before it becomes unstable (Cuk & Stewart, 2012).

The EBM self-consistently computes climate parameters over the entire orbital cycle around the star. It uses an explicit forward marching numerical scheme with a constant time step (i.e. a fraction of a day). Planetary and stellar parameters, including star type, semi-major axis, stellar insolation, orbital eccentricity, obliquity, and rotation rate can all be varied. The model can perform both mean-averaged and seasonally-averaged computations, reaching convergence when the annual-averaged mean surface temperature variation following each orbit falls below a threshold value (0.01 K).

*3.4 Climate modeling procedures*

We use our EBM to compute the circumstellar zone in which clathrate-rich subpolar ice could form on ocean world planets. We calculate this ice cap zone for the Sun, K2 M0, M3, M5, and M8 stars, which corresponds to the following stellar luminosities: 1 $L_{sun}$, 0.29 $L_{sun}$, 0.08 $L_{sun}$, 0.025 $L_{sun}$, 0.011 $L_{sun}$, and 0.0009 $L_{sun}$, respectively. The corresponding stellar masses in terms of solar masses ($M_{sun}$) are estimated to be: 1.5, 1, 0.79, 0.5, 0.15, and 0.08. Even though spectral classes less massive than 0.3 solar masses (which we take as 3400 K) are likely be tidally-locked (Leconte et al. 2015), we include the entire spectrum of M-stars because they may still be rapidly rotating under extreme circumstances (pre- to post-accretion, as with the early Earth). Following Levi et al. (2017), we assume that a huge reservoir of water is locked in the water ice mantles of these water worlds so that any water losses that M-star planets may suffer during an early period of intense atmospheric escape (e.g. Ramirez & Kaltenegger, 2014; Luger & Barnes, 2015; Tian & Ida, 2015) would be easily replenished. Worlds located around G – K stars are located far enough away from their stars to avoid the worst of pre-main-sequence water losses.

As has been shown (e.g Joshi and Haberle, 2012; Shields et al. 2013), the stellar energy distribution (SED) determines the fraction of visible and near-infrared radiation reflected away by surface ice. Using Bt-Settl stellar spectra data (Allard, 2003; 2007), we determine that ~53.3%, 34%, 12.7%, 9%, 5.4%, and 0.06% of the reflected radiation occurs at visible wavelengths for the Sun, K2, M0, M3, M5, and M8 stars, respectively. These values determine the surface albedo of the sea ice. Following Haqq-Misra et al. (2016), we assume that $CO_2$ condenses on to the surface and has a characteristic albedo of 0.35.

We wish to compare against the HZ as originally defined by Kasting et al. (1993) and assume all atmospheres are fully-saturated and contain 1 bar of $N_2$. For each grain size (100, 200, 300 microns) we find the stellar distance that gives the appropriate $pCO_2$ value at the inner edge of the ice cap zone and another distance corresponding with the higher $pCO_2$ value calculated for the outer edge (see section 3.1). For the inner edge we find the stellar distance that insures the average annual subpolar temperatures discussed in Section 3.1. All computed average subtropical temperatures were at least 265 K and no higher than ~ 335 K. The latter is lower than the maximum temperatures that life can sustain on the Earth (~390 K; Clarke, 2014). This is also lower than that required to trigger a moist greenhouse on a planet with an Earth-like water inventory (340 K). We note that at the highest rotation rates (10x) near the inner edge, equatorial temperatures do exceed 340 K (see Results) even though mean surface temperatures are below this value. Nevertheless, our ocean worlds are so water-rich that any water losses would not be major concerns in any case.

We assume an Earth-like 55% cloud cover at the inner edge because subtropical temperatures are often well above 273 K, which suggests that robust convection and cloud formation should occur at those latitudes. We use a procedure similar to that employed in Haqq-Misra et al. (2016) and assume that radiative forcing (negative) from clouds is ~6.9 W/m$^2$, which is equal to that the model needs to achieve a mean surface temperature of 288 K for the Earth. At the outer edge, temperatures are too low for water vapor clouds to persist (following Kasting et al. 1993) so we assume no radiatively-active water clouds there. We vary $\Omega$ from $1\Omega o$ - $10\Omega o$, with the latter corresponding to the fastest rotation rate possible for terrestrial planets before they destabilize (Cuk & Stewart, 2012). We assume an Earth-like gravity for a 2 Earth-mass planet with 35% of its mass in water (Levi et al., 2014). This is consistent with a terrestrial planet that has a bulk density of ~ 3.9 g/cm$^3$ (similar to that for Mars) and a radius of $1.41 R_{earth}$. An Earth-like obliquity of 23.5 degrees is assumed. Orbits are assumed to be circular. Land coverage is 0%.

## 4. RESULTS

We find that the ice cap zone is located within the classical $CO_2$-$H_2O$ habitable zone (Kasting et al. 1993). The incident stellar fluxes and corresponding edges of the classical and new ice cap zone for G-M-stars are shown in Fig. 4. Unlike the classical HZ, which is based on atmospheric modeling calculations (e.g. Kasting et al. 1993;

Kopparapu et al. 2013), we also present an alternative HZ limit that is not based on atmospheric models, but on empirical observations of the solar system (Fig. 4). The inner edge of this empirical HZ is defined by the stellar flux received by Venus after which standing water ceased to exist on its surface (~1 Gyr ago), equivalent to an amount equal to 1.77 times that received by the Earth or an effective stellar flux ($S_{eff}$) of 1.77 (Kasting et al., 1993). The outer edge is defined by the stellar flux that Mars received at the time that it may have had stable water on its surface (about 3.8 Gyr ago), which corresponds to an $S_{eff}$ = 0.32 (ibid).

We compute the $S_{eff}$ values for the inner and outer edges of the classical HZ by using the standard formulation from our previous work (eqn. 11) (e.g. Kopparapu et al., 2013; Ramirez & Kaltenegger, 2017):

$$S_{eff} = S_{eff(sun)} + aT_* + bT_*^2 + cT_*^3 + dT_*^4 \quad (11)$$

where $T_*$ is equal to $T_{eff}$ – 5780K, $S_{eff(sun)}$ are the fluxes computed for our solar system, and $a,b,c,d$ are constants, following Kopparapu et al. (2013). The corresponding orbital distances can then be computed with eqn. 12 (e.g. Kasting et al., 1993):

$$d = \sqrt{\frac{L/L_{sun}}{S_{EFF}}} \quad (12)$$

Here, $L/L_{sun}$ is the stellar luminosity in solar units and $d$ is the orbital distance in AU. The corresponding $S_{eff}$ values for the inner and outer edges of the ice cap zone are determined from solving the above expression for $S_{eff}$ (Fig. 4):

Although not as wide as the classical or empirical HZ boundaries, the ice cap zone is a significantly wide region worthy of follow-up observations. Earth-like rotation rates ($\Omega = \Omega o$) fail to provide the equator-pole temperature gradients necessary to support the cold subpolar and warm subtropical regions needed for a given atmospheric $CO_2$ level. In such cases, we were not able to find a distinct ice cap region (Fig. 5). Ice cap regions were found for $\Omega$ equal to or exceeding ~$3\Omega o$. At $\Omega = 3.3\Omega o$, the ice cap region using the smallest ice Ih grain size (100 microns) is 1.37 – 1.65 AU, 0.79 - 0.954 AU, 0.452 – 0.528 AU, 0.256 – 0.306 AU, 0.171 – 0.206 AU, and 0.0502 – 0.0617 AU for G2, K2, M0, M3, M5, and M8 stars, respectively. The width of this ice cap zone decreases at the larger grain size (300 microns) by ~45%, 45%, 49%, 42%, 36%, and 42% for the various stars (Fig. 4a).

The ice cap zone is even wider at the highest rotation rate ($10\Omega o$). The smallest assumed grain size (100 microns) for this case yields the widest possible ice cap region (Fig. 4b): 1.23 – 1.65 AU, 0.69 - 0.954 AU, 0.38 – 0.528 AU, 0.219 – 0.308 AU, 0.146 - 0.206 AU, and 0.0428 -0.0617 AU, for the various stars, respectively. This zone decreases in width by 17%, 9%, 22%, 19%, 25%, and 17%, respectively, at the largest grain size considered (300 microns). If such high rotation rates seem strange, we note that without the stabilizing influence of the Moon, Earth would be rotating much faster than it is today (e.g. Mardlin and Lin, 2002).

We also show typical mean-averaged temperature distributions at and near the inner edge for the 200 micron $3\Omega o$ and $10\Omega o$ rotation rate case (Fig. 6). We summarize inner edge boundary distances (in AU) for the ice cap zone for the various grain sizes and rotation rates, assuming the stellar luminosity values mentioned in Section 3.4 (Table 1).

The outer edge boundaries for the ice cap zone, as mentioned earlier, remain unchanged from one grain size to the other.

However, we find that these values are slightly less than for the classical HZ (Fig. 4) because the latter calculations assumed a surface albedo consistent with an ice-free surface, which slightly overestimates the absorption and the HZ width.

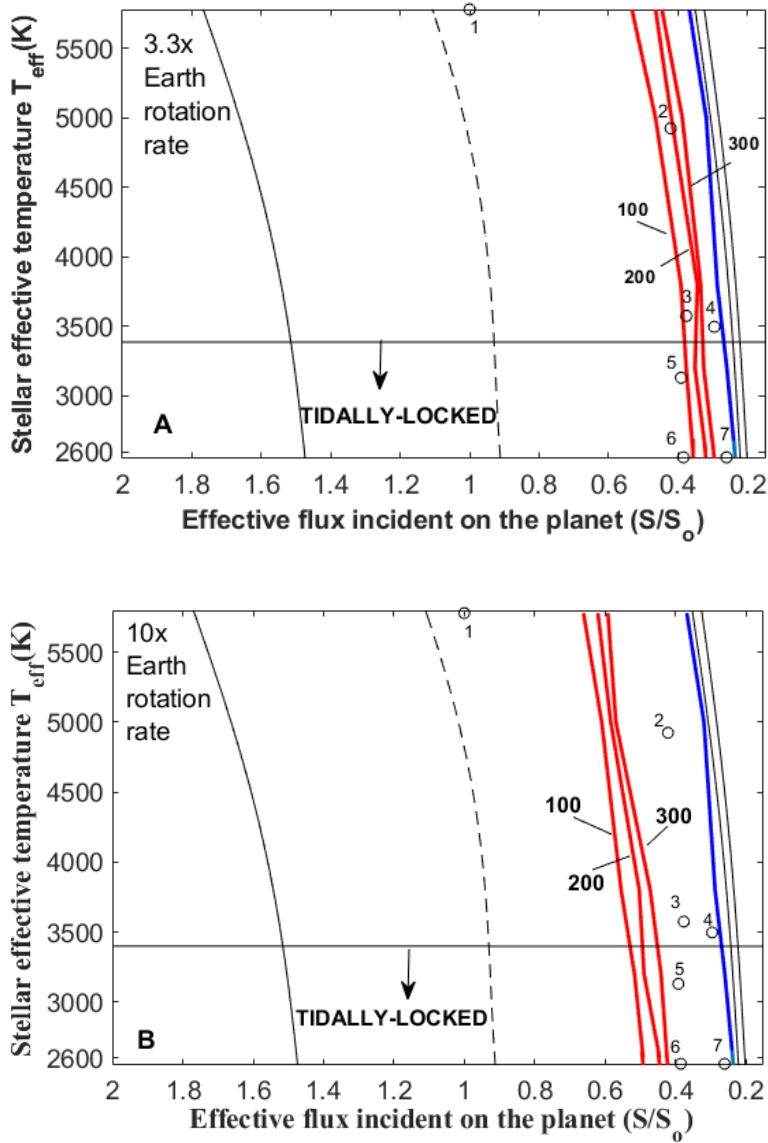

**Figure 4:** Effective stellar temperature versus incident stellar flux ($S_{eff}$) for the empirical (thin black lines), classical (dashed black lines), and the inner edge of the ice cap zone for the 3 different (100, 200, 300 micron) grain sizes (thick red lines) assuming a) 7.2 and b) 2.4 hour rotation rates with corresponding outer edge (thick blue line). Three confirmed planets are located within the ice cap zone. The labeled planets are as follows: 1:Earth, 2:Kepler-62f, 3:Kepler-1229b, 4: GL-581d 5: LHS-1140b, 6: TRAPPIST1-f, 7: TRAPPIST1-g. Planets orbiting stars cooler than ~ 3,400 K are likely to be tidally-locked.

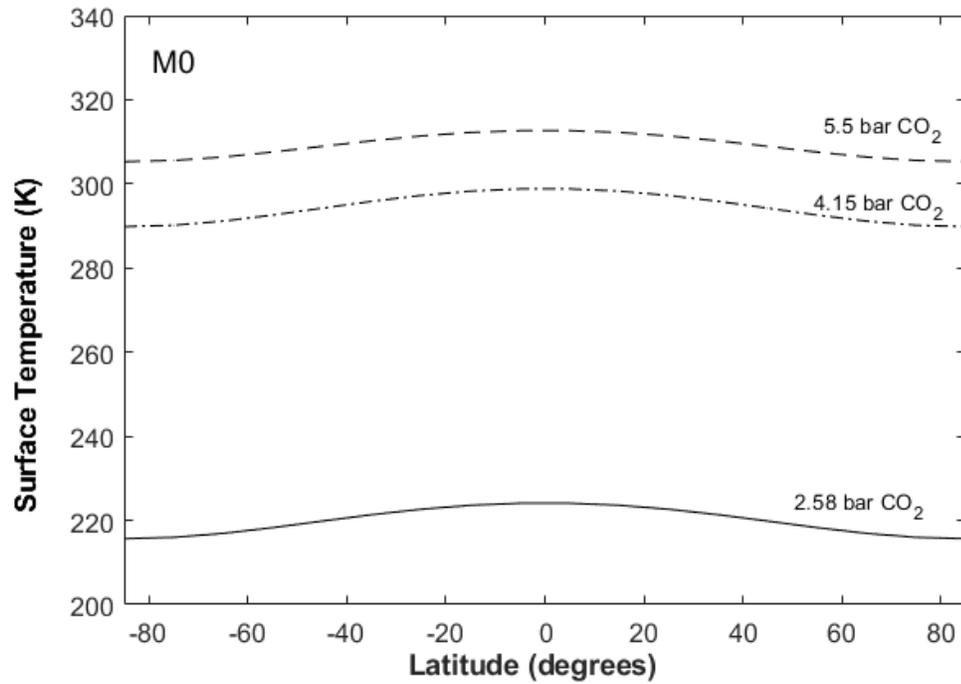

**Figure 5:** Mean annual surface temperature distributions for the 2.58, 4.15, and 5.5 bar $CO_2$ scenarios at a distance of 0.475 AU from the M0 star for $\Omega = \Omega o$. Temperature gradients (from equator to pole) are 8, 9, and 7 K, respectively, All planets have 1 bar $N_2$.

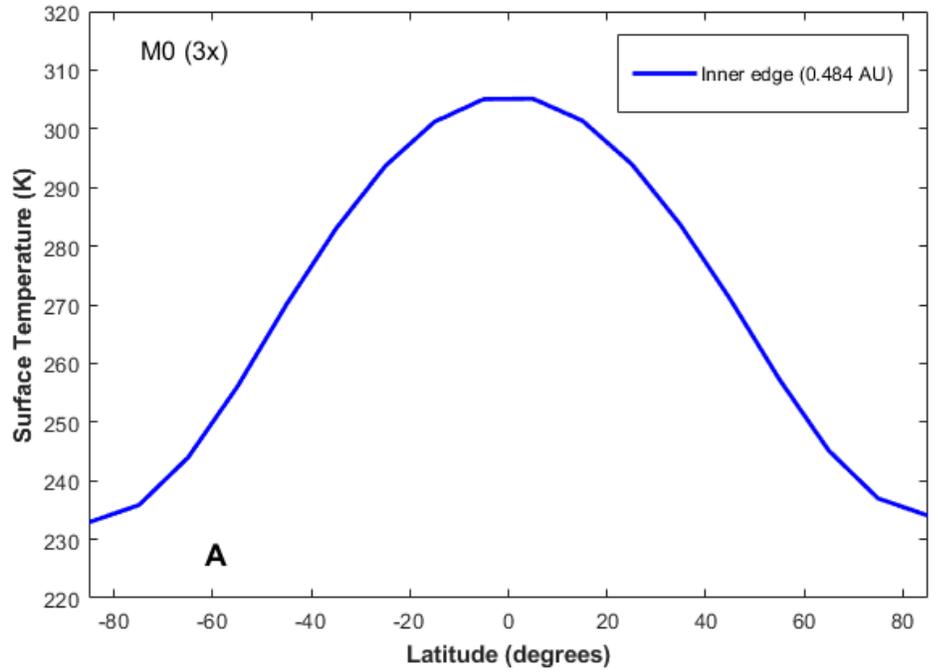

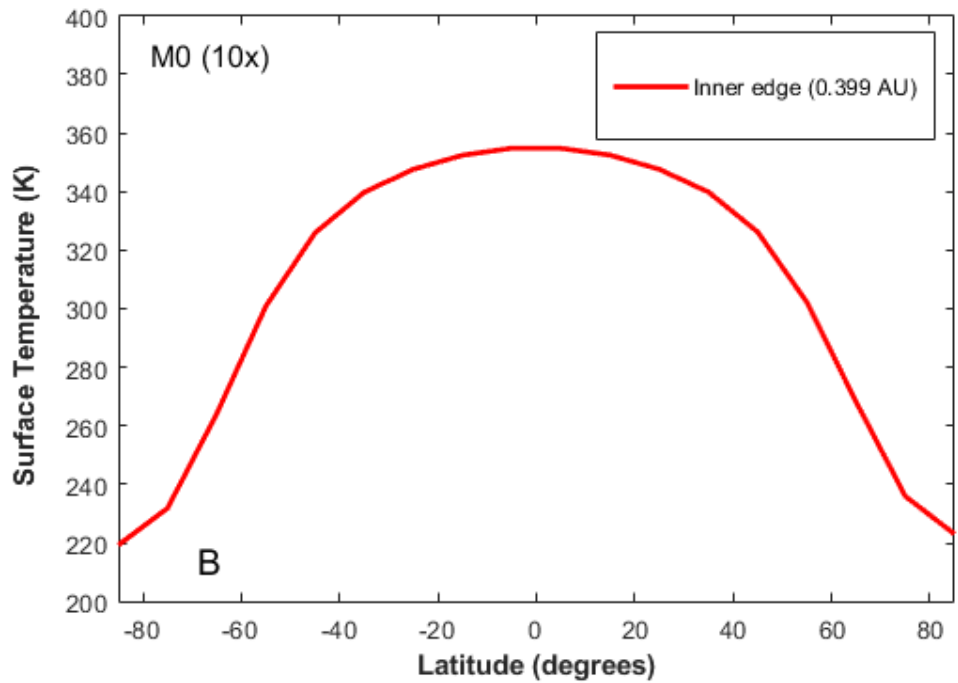

**Figure 6**: Mean annual surface temperature distributions for the 200 micron case (4.15 bar $CO_2$) at the inner edge for $\Omega =$ a) $3\Omega o$ and b) $10\Omega o$.

**Table 1**: Inner edge boundaries (in AU) for different rotation rates (Ω) and grain sizes (in microns)

| Stellar Class | 3Ωo | | | 10Ωo | | |
|---|---|---|---|---|---|---|
| | 100 μ | 200 μ | 300 μ | 100 μ | 200 μ | 300 μ |
| Sun | 1.37 | 1.467 | 1.497 | 1.23 | 1.27 | 1.3 |
| K2 | 0.79 | 0.832 | 0.865 | 0.69 | 0.705 | 0.714 |
| M0 | 0.452 | 0.482 | 0.489 | 0.38 | 0.399 | 0.412 |
| M3 | 0.256 | 0.268 | 0.277 | 0.219 | 0.226 | 0.236 |
| M5 | 0.171 | 0.177 | 0.1837 | 0.146 | 0.15 | 0.158 |
| M8 | 0.0502 | 0.053 | 0.055 | 0.0428 | 0.045 | 0.0461 |

## 5. DISCUSSION

*5.1 EBM Critique and Justification*

It is sometimes argued that the diffusion parameterizations that many EBMs use to describe equator-to-pole dynamic transport (e.g., eq. 6) oversimplify dynamical processes. Although this is a common critique against EBMs, we believe this argument is overstated because the errors in these aspects are dwarfed by other uncertainties regarding the exoplanets themselves (including cloud cover, cloud forcing, mass, atmospheric composition, rotation rate etc.), which would be problematic for any model to ascertain without detailed observations. For example, even though surface temperatures towards higher latitudes across the tropics decrease more rapidly in our EBM than what Earth observations suggest, these differences are relatively small (~1 to 3K) (Lindzen and Farrell, 1977) (Fig. 7). Moreover, EBMs tend to transition into an icy state somewhat more abruptly than GCMs do (Shields et al. (2013), but in spite of this, as our Fig. 7 shows, the northern hemispheric equator-pole temperature gradients in the EBM are similar to those observed for the Earth. Although observations suggest that the southern hemispheric pole is several degrees colder than what our model would predict, large topographic variations characterize that hemisphere. However, topographic features ought to be highly subdued on planets whose outer condensed layer is a global deep ocean, especially since the maximum sea ice (enriched in $CO_2$ clathrate) thickness is only ~2 m before sinking (Levi et al. 2017). We note that the best agreement between the EBM and Earth occurs in the northern hemisphere, where topographical variations are smallest (Sellers, 1969). This suggests that our model is appropriate for the flat and relatively homogeneous ocean worlds we assess in this study (as we argued in Section 3.3). In comparison, the mean surface temperature of the north pole in GCMs that lack ocean heat transport is too cold by ~20 K (e.g Shields et al. 2013). Our model also obtains a mean surface temperature of 288 K and a planetary albedo of ~0.3 for the Earth, which gives further confidence in its extrapolation to exoplanetary atmospheres.

Also, idealized GCM simulations (lacking ocean heat transport and/or employing simplified radiative transfer) of the Earth predict that latitudinal temperature gradients should increase at higher rotation rates (e.g. Jenkins, 1993; Charnay et al. 2013; Kaspi and Showman, 2015), which is broadly consistent with the diffusion parameterization used in many advanced

EBMs (eq. 6). Although cloud behavior changes with the rotation rate in those models (whereas we keep our cloud feedback constant), such results are consistent with our analysis here. The latter idealized GCM study showed that gradients continually increased from an Earth-like rotation rate to 8x that of Earth (Kaspi and Showman, 2015). That said, we do not know how well our diffusion parameterization agrees with full atmospheric-ocean GCM simulations at very high rotation rates for these high-$CO_2$ ocean world atmospheres. However, we expect that such GCMs with proper ocean heat transport should exhibit even better agreement with EBM simulations because ocean transport is implicit in our diffusion parameterization. A future study could assess this possibility.

Nevertheless, advanced EBMs continue and will continue to be successfully used for many planetary and exoplanetary applications (e.g. North & Coakley (1979), Williams and Kasting (1997), Caldeira & Kasting (1992), Vladilo et al. (2013;2015), Batalha et al. (2016); Forgan (2016), and Haqq-Misra et al. (2016). They are consistently able to obtain bulk trends that are consistent with observations and complex models in spite of their simplifications. Our EBM is no different in this regard. It is our hope that our study motivates future GCM modelers to assess how well predictions from such diffusion parameterizations agree with fully-coupled atmosphere-ocean model results in this high rotation rate regime for thick $CO_2$ atmospheres on ocean worlds. Such efforts would yield improved parameterizations for EBMs too. GCMs and simpler models (EBMs and single-column climate models) are both necessary to understand planetary climates (Jenkins, 1993; Pierrehumbert et al. 2007).

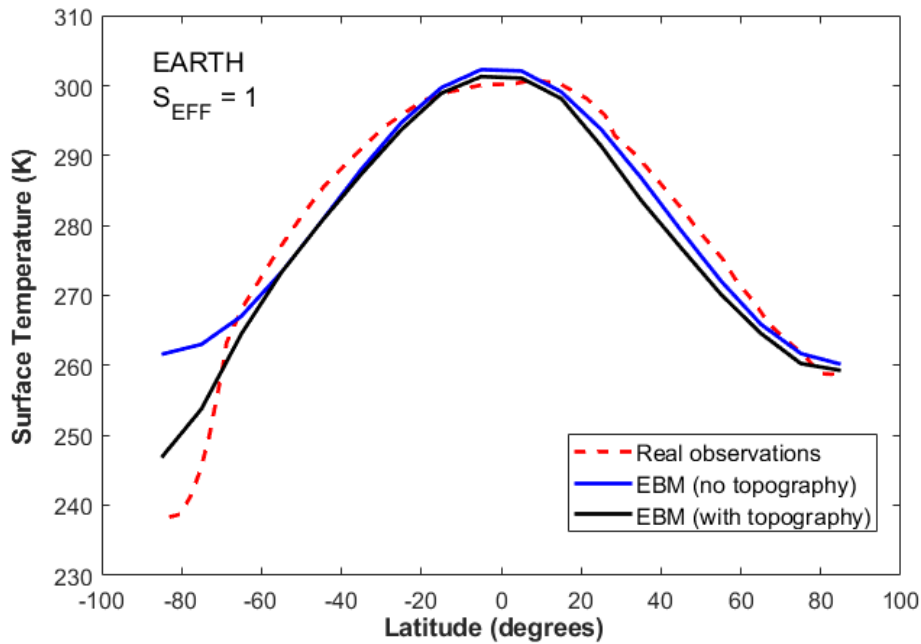

**Figure 7:** Mean annual surface temperature as a function of latitude for the present Earth comparing our model without topography and with topography against real observations. In the model case with topography, surface temperatures were assumed to decrease with elevation (zonally-averaged data from Sellers, 1969) at a characteristic lapse rate for the Earth (6.5 K/km).

*5.2 Sensitivity Study: Weak Thermal Gradient Approximation on Equator – Pole Gradients*

The Levi et al. (2017) mechanism operates under large equator-pole gradients. We wrote our own version of the weak thermal gradient (WTG) approximation (e.g. Pierrehumbert, 2010) to assess whether such large gradients may also be possible on tidally-locked worlds. To clarify, the WTG approximation only suggests that temperature gradients within the free atmosphere should be small, as expected for tidally-locked planets (e.g. Mills and Abbot, 2013). However, surface temperature gradients can still be quite large. We found that equator-pole surface temperature gradients in the most optically thin (favorable) scenario (3.58 bar, including 1 bar $N_2$) were only ~7 K, even assuming an ocean wind speed (U) of 5 m/s, which is the lowest considered by Levi et al. (2017). Subsequently, these gradients are slightly smaller than those computed by the EBM for the Earth-rotation cases (Fig. 5). Surface temperature gradients ($\Delta T$) approaching those necessary (> ~30 K) would only be possible at U ~ < 1 m/s, which is unrealistically low to support such large surface temperature variations, especially since U $\alpha$ $\Delta T^{2/3}$ (Cullum et al. 2014). Thus, this is consistent with our result that such equator-pole temperature gradients cannot be achieved if planets are not rotating rather quickly, as we have argued throughout our analysis.

*5.3 Evidence of ice cap zone for observations*

Our computed boundaries in Fig. 4 indicate that at least 3 currently confirmed planets (Kepler 62-f, GL-581d, Kepler-1229b), thought to be located in the classical HZ, are also in our ice cap zone. Our ice cap zone for M-stars is nominally valid for stars hotter than ~3400 K (~M3) because planets orbiting even cooler stars would likely be tidally-locked (Leconte et al. (2015) However, it is possible that planets orbiting stars cooler than ~3400 K may experience a brief period of rapid rotation post-accretion before undergoing synchronous rotation. During this brief period, these planets may experience numerous freeze-thaw cycles, depending on the time-scale for tidal locking. If the latter is long enough, material of biological interest may be produced.

The number of potentially habitable ocean worlds should correspondingly increase as upcoming missions, including the *James Webb Space Telescope* (JWST) and the *Transiting Exoplanet Survey Satellite* (TESS), find many new targets around low-mass stars. The C/O ratio is a key atmospheric quantity that JWST would infer from the atmospheric composition (e.g. Greene et al., 2016). If we assume that other greenhouse gas abundances are negligible and $CO_2$ is relatively well-mixed in the atmosphere, then column abundances(N) of $CO_2$ and $H_2O$ (N = P/g) can be used to readily derive the C/O ratio. For both edges of the ice cap zone, these ratios are ~ 0.48 – 0.5, assuming a range of representative mean surface temperatures: ~270 – 340 K for the inner edge and 250 K for the outer edge. All of our atmospheres are dominated by $CO_2$, resulting in a relatively narrow range of predicted C/O ratios. Rotation periods for exoplanets are difficult to ascertain observationally. This work shows that for ocean worlds located within the ice cap zone, the atmospheric C/O ratio can place an upper boundary on the rotation period (Fig. 4).

Also, some ocean worlds may be inferred directly from mass-radius observations if they consist of enough water by mass. For instance, a 2-Earth mass planet with 50% water (by total mass) has a bulk density of ~ 3 g/cm$^3$, considerably lower than for the Earth (~5.5 g/cm$^3$) or Mars (~3.9 g/cm$^3$). Such planets would have low atmospheric scale heights and observations would easily distinguish them from planets with massive H/He envelopes. Also, their low bulk densities should be considerably lower than those for $CO_2$-rich terrestrial planets with Earth-like water inventories associated with the classical HZ. Therefore, the bulk density is an initial filter, helping to pinpoint ocean worlds within the ice cap zone. The second filter would be the C/O ratio helping to pinpoint those planets that are more likely to be of biological interest.

*5.4 The importance of rotation rate*

Our results show that a different ice cap zone can be defined as a function of the planetary rotation rate and the assumed particle size (Table I). Regardless of rotation rate (10$\Omega o$ or 3.3$\Omega o$), at least 3 confirmed planets are located within it (Fig. 4). Although our study is the first to show the importance of rotation rate in $CO_2$-rich atmospheres near the outer edge of the classical HZ, its importance has also been demonstrated in studies of early Venus (e.g. Yang et al. 2014; Way et al 2016). These studies effectively suggest that the inner edge of the classical HZ may move to shorter distances if it is assumed that planets can rotate much more slowly than the Earth.

*5.5. Analysis of mechanism, caveats, and future work*

Our work here is meant to test the viability of the Levi et al. (2017) mechanism and to derive inner and outer edges for potentially habitable ocean worlds around various stars. We believe we have shown that this hypothesis may be plausible for G2 – M3 stars although our mechanism does not seem very plausible for stars cooler than ~3400 K.

We have provided limits for the ice cap zone that are consistent with the atmospheric conditions assumed. However, we note a few caveats that could be assessed in future work. The biggest uncertainty in these and all other HZ calculations has arguably been clouds. Clouds would conceivably affect the latitudinal temperature-distribution relative to our case in which we keep cloud coverage constant. This could be studied in future work. Although the relative importance of clouds would very much depend on the specific conditions of a given planet, future work could assess different scenarios (e.g. cloud coverage, distribution) and gauge their effects on the boundaries. We also assumed a fixed surface albedo for $CO_2$ ice. However, this would conceivably be a function of local variables, including ice thickness, atmospheric pressures, and temperatures. Another big uncertainty is the dynamics of sea-ice and how such ice near the equator would operate at temperatures near the freezing point of water. This would affect the temperature below which global glaciation ensues. However, sensitivity studies (not shown) revealed that a threshold temperature of 273 K instead of 265K would only have a small effect on the outer edge distance (~ 1%). A final uncertainty is the ice-albedo feedback, which is parameterized differently in different models (e.g. Curry et al., 2001), and ours is no different in this regard. Small differences in this may slightly affect the computed inner and outer edges. Given the uncertainties in all such calculations, improved observations of exoplanetary atmospheres would increase our understanding and yield substantial improvements to all (i.e. 1-D and 3-D) exoplanet climate models.

## 6. CONCLUSION

We have found that a circumstellar region around G-M stars in which the $CO_2$ cycling mechanism of Levi et al. (2017) operates on ocean worlds could exist and that habitable conditions may be possible for planets that rotate at least ~ 3 times faster than the Earth. These high rotation rates generate the required equator-pole temperature gradients necessary to sustain both cold subpolar and warm subtropical regions in these dense (> ~ 2.5 bar) $CO_2$ atmospheres. This is important for establishing freeze-thaw cycles, needed to help life evolve in diluted ocean worlds. The inner edge of this ice cap zone is moderately dependent on ice grain size whereas the outer edge is relatively insensitive to this parameter. The widest ice cap zone, assuming 100 micron ice Ih grains and a rotation rate of ~2.4 hours, is ~1.23 - 1.65, 0.69 - 0.954, 0.38 – 0.528 AU, 0.219 – 0.308 AU, 0.146 - 0.206 AU, and 0.0428 - 0.0617 AU, for G2, K2, M0, M3, M5, and M8 stars, respectively. However, unless the planets are very young, we expect that planets orbiting stars cooler than a ~ M3 will be tidally-locked and so our mechanism would not apply to such planets. This zone decreases in width at larger ice Ih grain sizes and slower rotation rates. We predict C/O ratios for our atmospheres (~0.48 – 0.5) that can be verified by future missions, including JWST.

## ACKNOWLEDGEMENTS

We thank Mercedes Lopez-Morales and James F. Kasting for fruitful discussions regarding C/O ratios and JWST observations. We also acknowledge helpful conservations with Lisa Kaltenegger and Dimitar Sasselov. We also thank the anonymous referee for constructive comments which improved our manuscript. Both R.M.R and A.L. acknowledge support by the Simons Foundation (SCOL No. 290357, L.K.; SCOL No. 290360). R.M.R also acknowledges support by the Carl Sagan Institute and the Earth-Life Science Institute.

Ramirez, R. M., & Kaltenegger, L. (2017). A volcanic hydrogen habitable zone. The Astrophysical Journal Letters, 837(1), L4.

Ramirez, Ramses M. "A warmer and wetter solution for early Mars and the challenges with transient warming." Icarus 297 (2017): 71-82.

Schwartz, Alan, and M Goverde. 1982. "Acceleration of HCN Oligomerization by Formaldehyde and Related Compounds: Implications for Prebiotic Syntheses." *Journal of Molecular Evolution* 18: 351-353.

Seager, Sara. "Exoplanet habitability." *Science* 340.6132 (2013): 577-581.

Sellers, William D. "A global climatic model based on the energy balance of the earth-atmosphere system." *Journal of Applied Meteorology* 8.3 (1969): 392-400.

Shields, A. L., Meadows, V. S., Bitz, C. M., Pierrehumbert, R. T., Joshi, M. M., & Robinson, T. D. (2013). The effect of host star spectral energy distribution and ice-albedo feedback on the climate of extrasolar planets. Astrobiology, 13(8), 715-739.

Simpson, Fergus. "Bayesian evidence for the prevalence of waterworlds." *Monthly Notices of the Royal Astronomical Society* 468.3 (2017): 2803-2815.

Slack, Glen. 1980. "Thermal conductivity of ice." *Physical Review B* 22: 3065 .

Tian, Feng, and Shigeru Ida. "Water contents of Earth-mass planets around M dwarfs." Nature Geoscience 8.3 (2015): 177.

Vladilo, Giovanni, et al. "The habitable zone of Earth-like planets with different levels of atmospheric pressure." *The Astrophysical Journal* 767.1 (2013): 65.

Vladilo, Giovanni, et al. "Modeling the surface temperature of Earth-like planets." *The Astrophysical Journal* 804.1 (2015): 50.

Way, M. J., Del Genio, A. D., Kiang, N. Y., Sohl, L. E., Grinspoon, D. H., Aleinov, I., ... & Clune, T. (2016). Was Venus the first habitable world of our solar system?. *Geophysical Research Letters*, *43*(16), 8376-8383.

Williams, D. M., & Kasting, J. F. (1997). Habitable planets with high obliquities. Icarus, 129(1), 254-267.

Wordsworth, R., F. Forget, and Vincent Eymet. "Infrared collision-induced and far-line absorption in dense CO 2 atmospheres." *Icarus* 210.2 (2010): 992-997.

Wordsworth, Robin D., and Raymond T. Pierrehumbert. "Water loss from terrestrial planets with CO2-rich atmospheres." *The Astrophysical Journal* 778.2 (2013): 154.
30